\documentclass[12pt]{article}
\usepackage{epsf}

\newcommand{\be}{\begin{eqnarray}}

\newcommand{\ee}{\end{eqnarray}}

\newcommand {\la}{ \,\, 
 \vcenter{\hbox{$\buildrel{\displaystyle <}\over\sim$}} \,\,}

\newcommand{\MeV}{\hbox{MeV}}

\newcommand{\fm}{\hbox{fm}}
\newcommand{\km}{\hbox{km}}

\newcommand{\scrpt}[1]{{\hbox{\scriptsize #1}}}

\title{
        \begin{flushright}
        \normalsize
	TPI-MINN-97/28\\
	NUC-MINN-97/12-T\\
	UMN-TH-1612
        \end{flushright}
\bf     Hot Neutron Stars as a Source for Gamma Ray Bursts
        at Cosmological Distance Scales       
       }
\author{Henning Heiselberg \\
        {\small\it NORDITA,
        Blegdamsvej 17,
        Copenhagen DK } \\ 
        Sangyong Jeon, Larry McLerran and Hua-Bin Tang\\
        {\small\it Physics Department, University of Minnesota, 
        Minneapolis, MN 55455} \\
        }

\date{}

\begin{document}
\setcounter{page}{0}
\maketitle
\thispagestyle{empty}
\begin{center}
{\bf Abstract}\\
\end{center}
We discuss the possibility that the sources for gamma ray bursts are
hot neutron stars at cosmological distance scales.  
The temperature of such stars would be $T \sim 1\,\MeV$.  
Such hot stars can produce an electromagnetic blast wave 
provided that the ratio of baryon and photon numbers
 $N_{B}/N_{\gamma} \le 10^{-6}$.  
The typical time scale for such blasts, the total luminosity, 
and correlation of gamma ray energy with time of arrival are shown 
to be roughly consistent with observation.  
The spectrum of photons also appears to be consistent with known data.

\noindent 

\vfill \eject

\section{Introduction}

Gamma ray bursts (GRBs) have been the subject of much theoretical and
experimental study\cite{nemiroff,fishman,higdon}.  The source of
these bursts is the subject of much dispute.  If the source is near
our galaxy, then such objects might be neutron stars\cite{lamb}.
If on the other hand, such objects are at cosmological distance scales, then
the total energy released in photons is typical of supernova
explosions.  In this paper, we will assume that gamma ray bursts are
at cosmological distance scales.  

The typical energy and number of such energetic photons
are so large that a conventional supernova explosion
will not describe the bursts. In a supernova,
most photons are soft.  The neutrinos carry energies typical of one 
MeV scale, but the photon energy is degraded.   
Any initially energetic photons are produced
in an optically thick region.  The photons one sees are produced much
later than the time at which the energetic neutrinos and photons are 
produced, and are emitted after the supernova remnants have cooled.
  
At first sight, it would seem impossible that photons produced from
a hot source in the center of some violent explosive event could
maintain their energy.
In the seminal work by Paczynski\cite{paczynski} and by
Goodman\cite{goodman}, it was shown that under certain circumstances
this is indeed possible.
In this work, it was suggested that energetic photons 
emitted from some hot source
would expand hydrodynamically.  It is assumed that to a good
first approximation, one has an expanding gas of photons and perhaps
initially (depending upon the initial temperature)
electron-positron pairs in thermal equilibrium.
The ratio of baryons to photons, $N_B/N_{\gamma}$, is assumed to
be very small.

In this hydrodynamical expansion,
both the energy and entropy are conserved.  The energy per photon is
thus roughly constant.  (There may be an order one change due to 
the annihilation of electron-positron pairs as the hot plasma cools.)
Near the hot source, the photon gas is moving
outward slowly and the energy per photon is large, since the initial
temperature is assumed to be large.  
Far away from the source, the fluid is relativistic 
and has a large Lorentz $\gamma$ factor.
The local rest frame temperature is however small, of order
$T_s/\gamma$, where $T_s$ is the surface temperature.  
The local temperature
has been traded for the relativistic gamma factor of the fluid,
and this allows for the constant energy per photon.

In order that the energy per photon be roughly conserved, it is necessary that
the baryon to photon ratio, $N_B/N_{\gamma}$, be small.  
We will see that this typically requires $N_B/N_{\gamma} \le 10^{-6}$.  
This small ratio at first sight seems unnatural since any mechanism 
which produces the photons must almost certainly come from a baryon 
rich region.  If the ratio is larger, the energy of the gammas 
at the source is significantly degraded at decoupling from the moving fluid.

This picture of Paczynski and of Goodman has led to a blast wave
description
of the later stages of such an explosive process\cite{rees}.  It is
claimed that it has the correct properties to explain the
afterglow\cite{waxman}
which has been recently observed in the visible
spectrum\cite{paradjis}.
If the observations of this afterglow are correct, then some fraction
of gamma ray bursts must occur at cosmological distance scales.

In this conventional picture \cite{waxman}, 
a GRB is produced in two stages.  
First, a yet unspecified event generates optically thick plasma by
radiating a huge amount of heat in a few tens of seconds.
Eventually, the energy of this relativistically expanding fireball 
is all converted to the proton kinetic energy. 
In the next stage, the proton kinetic energy dissipates 
producing GRBs either by internal collision and/or
by the collisions with interstellar medium.

Our picture here is somewhat different from the above conventional one.
In our scenario, the initial photons are the direct source of GRBs. 
Hence, we require that the protons do not
absorb a large portion of the initial photons.

Regardless of the difference, the conventional picture and ours share 
a common theme -- the blast wave. 
In this paper, we address the issue of how such a blast wave is
generated.
Since the total energy in such a blast wave is so large, a source such
as a neutron star or the collision of neutron stars would appear to be
a likely candidate.  
We will accept this hypothesis here and then see whether the gross
properties of gamma ray bursts can be explained in such a picture.

There have been a multitude of suggestions for neutron-star mechanisms
for generating gamma ray bursts at cosmological distance scale.  The
observed energy per photon and the total luminosity require an energy
release per proton of about $E/N_{B} \sim 1-10\,\MeV$.  The classes of
models
which could generate such an energy release might be the following:
\begin{enumerate}

\item{\bf 
Neutron (Strange) Star Collisions or Collapsing Neutron Star Binaries}

In neutron star collisions or the collapse of neutron star binaries,
some of the gravitational binding energy is released as
heat\cite{schramm}.
If the stars collapse to a black hole, the gravitational binding
energy of a particle is its rest mass.  Only a small fraction of this
energy needs be converted into radiation.  
Whether or not the time scales for such collapse are short 
enough\cite{swesty} is not certain, and whether some hot emitting region 
can live long enough outside the
event horizon is not clear.

Another possibility is a collision of two small neutron stars producing 
not a black hole but a bigger neutron star.  The time scale of this may be
just about what we need for a GRB. 

Yet another possibility is collisions or a collapse involving strange
stars.  
A strange star as a GRB source may have an additional benefit of  
possessing less volatile surface than that of a neutron star. 

\item{\bf Nucleation of a Black Hole Inside a Neutron Star}
 
It might be possible that a black hole is spontaneously nucleated
inside a neutron star\cite{voloshin}.  If so, heat would be released
in the core of the star which would heat the surface layers and serve
as an emitting region.  Whether this emitting region could live long
enough to describe gamma ray bursters is not clear \cite{Ruffert}.

\item{\bf Phase Transition of Matter in the Core of a Neutron Star}

A phase transition between different phases of matter might take
place in the core of a neutron star \cite{baym,olinto,cheng}.
We will not be concerned with the type of phase transition here except
to note that the natural energy scale in the core is of order 
$E/N_B \sim (10 - 100\,\MeV)$ per nucleon 
which when spread out over the entire
star should be large enough to describe gamma ray bursts.  Moreover,
the star will maintain its integrity after the phase transition since
the energy release is small compared to the gravitational binding
energies of nucleons in the star.

\end{enumerate}

The list above may not be complete.  Certainly some or perhaps all of
the mechanisms above may not describe gamma ray bursts and more
detailed computations within detailed models would be necessary
to obtain better resolution of the correctness of the various
hypothesis.

We will in this paper make a few simple generic assumptions about the nature of
the source of gamma ray bursts and explore the generic consequences
of these assumptions.  We specifically assume that the source region
is of size and baryon density typical of a neutron star.  We assume
that somehow this region maintains its integrity over the lifetime
of the gamma ray burst.  We assume the typical energy per
nucleon is typically of order $1 - 10\,\MeV$ per nucleon.

With these assumptions, we will make a hot neutron star model of the
gamma ray burst.  
For this model to work, we must show that 
\begin{enumerate}
\item
The time scale of the evolution of the burst is $\sim 1\,\sec$.
\item
The surface of the neutron star
heats up quickly after the initial generation of heat.
\item
The surface temperature ($\sim 1\,\MeV$), when it is established, 
is not too much smaller than the core temperature.
\item
The baryon contamination of the gamma-sphere is small
enough so that at the decoupling, a large number of energetic photons 
originating from the surface survive. 
\end{enumerate}

We will find that the natural time scales for the gamma
ray burst work out correctly as a consequence of this model.  
We also crudely compute the time evolution of the luminosity 
and energy per photon and argue that this is generically correct.  
The differential energy spectrum
of emitted photons also appears to be qualitatively correct and does not
fall as rapidly with decreasing energy as the fireball model of 
Goodman \cite{goodman}.

The surface temperature of a hot neutron star is estimated by assuming
that the hot core generates thermal neutrinos via modified Urca process,
and the surface radiates heat by the neutrino pair bremsstrahlung. 
We first show that the typical neutrino mean free path is so long 
that a neutrino interacts at most once with an electron inside the star.  
Hence the neutrinos from the hot core transport heat 
to the surface with the speed of light.  
Assuming that the dominant cooling process near the surface is 
the neutrino pair bremsstrahlung,
we will find that the surface temperature is not much different 
from the core temperature.

The issue of $N_B/N_\gamma$ is more problematic.
We have several suggestions which may allow for a sufficiently small
$N_B/N_\gamma$ so that our model works:
(i)~ a strong magnetic field near the neutron star surface,
(ii)~the formation of super-heated Coulomb crystal near 
the surface, or
(iii)~the nucleon binding energy of a strange star.
Certainly, the Coulomb crystal should exist in a cold
neutron star, and strong magnetic field is believed to exist at pulsars. 

The physical picture we generate as a consequence of this model
seems to be consistent.  The region from which the photons
decouple from the hot expanding electromagnetic plasma is large
compared to a typical neutron star size.   Therefore the details of
the shape of the emitting region which generates the electromagnetic
plasma is not important.  It need only be some region of the right
dimensions, temperature and density.  The calculation of the 
baryon to photon ratio
 will involve understanding the local emission from a hot
surface.
Again, the details of the shape and dynamics of the emitting region
will not be important.
 
\section{The Hot Neutron Star}

We assume that a large amount of heat is deposited into
the core of a neutron star, which gives an energy of about 10 MeV per nucleon.
We assume the time scale for this deposition is much less than the 
characteristic time scale for a gamma ray burst which we take to be of
order $t_{\scrpt{burst}} \sim 1\,\sec$.  

The first question we have to ask is
how disruptive is this deposition process, and the thermal emission
which follows the deposition, to the star.  In an ordinary star,
such energy deposition would be catastrophic.  There would be an
explosion
and a large fraction of the star would be blown away.  This is not the
case
for a neutron star however.  The typical scale in such a star is set
by Fermi energies which are typically hundreds of MeV.  The
gravitational
binding energy of a nucleon is also hundreds of MeV.  The energy deposited
is small compared to these scales, and we therefore do not expect a
catastrophic disruption of the star.

On the other hand, near the surface of the star, there will be emission
of some of the baryons from the star.  We will return to this problem later.

The initial energy deposited into the star can be spread throughout the
star by several mechanisms.  Shock fronts and burning fronts can be 
generated.  These fronts typically have relativistic velocities 
in a neutron star, and since the size is small $R \sim 10\,\km$, 
the time scale for spread of this energy ($\sim 10^{-4}\,\sec$)
is also much less than a second.

In addition to collective mechanism for transfer of heat, 
neutrinos also carry and deposit heat throughout the neutron star.  
The scattering mean free path for a neutrino is
\be
\lambda_{\scrpt{scat}} = {1 \over {\sigma_{\nu e}\, \rho_e\, (E_\nu/\mu_e)}}
\;,
\ee
where $\sigma_{\nu e}$ is the neutrino-electron scattering cross
section, $\rho_e$ is the electron number density, and $\mu_e$ is the
electron chemical potential.  The factor $(E_\nu/\mu_e)$ accounts for
the Pauli blocking. 
The neutrino-electron total cross section is given by
$\sigma_{\nu e} \sim (E_{\nu}^2/1\,\MeV^2)10^{-19}\,\fm^2$\cite{shapiro}.
Since $E_\nu \sim 3T_c$, where $T_c$ is the core temperature,
this gives a typical mean free path  
\be
\lambda_{\scrpt{scat}} \sim 
10^3\,\km\ 
\left({1\,\MeV \over T_c}\right)^3
\left({\rho_\scrpt{nuc} \over Y_e \rho_B} \right)^{2/3}
\;,
\label{eq:nu_mfp}
\ee
where $\rho_{\scrpt{nuc}} = 0.17\,\fm^{-3}$ is the nuclear matter
density and $Y_e \equiv \rho_e/\rho_B$ is the electron-baryon ratio. 
For most neutrinos, the mean free path is much longer
than the size of the neutron star.

One might suspect that since the neutrino mean free path is so long, the
surface temperature could not be maintained after the initial rise to
$\sim 1\,\MeV$.  This is not so.
The amount of energy flowing out of the core can be enormous
since the rate is proportional to $T_c^8$ (modified Urca process).
Hence, $T_c \sim 5\,\MeV$ can easily maintain the surface temperature of
$1\,\MeV$ even if the fraction of energy deposited is small.  

To see that this deposit is big enough to maintain the surface temperature
consider a steady state maintained by the modified
Urca process near the core and direct Urca process near the surface.
The hot core puts out neutrino energy with the rate per volume \cite{shapiro}
\be
{d^2E \over  dtdV} = \kappa_\scrpt{Urca} T^8
\;,
\ee
where 
\be
\kappa_\scrpt{Urca} 
\sim
10^{-4} \,\MeV^{-7} \fm^{-3} \sec^{-1} 
\left( {\rho_c \over \rho_\scrpt{nuc}} \right)^{2/3}
\;.
\ee
Assuming uniform temperature within a core radius $R_c$, 
the total energy output per unit time is 
\be
\left. {dE\over dt} \right|_{R_c}
= 
\kappa_\scrpt{Urca} T_c^8\, {4\pi\over 3}R_c^3
\;.
\label{eq:CoreLoss}
\ee

Since the neutrino mean free path is typically much longer than the
neutron star radius $R$, most of neutrinos originating 
from this hot core reaches $R$ unscathed.  
The survival probability of these neutrinos crossing a radial distance $dR$ 
near $R$ is given by 
\be
{\cal P} = 1 - {dR\over \lambda_{e \nu}} \ ,
\ee
where 
the neutrino mean free path is
\be
\lambda_{\nu e} = 1/\sigma_{\nu e}\rho_e
\;.
\ee
This differs from Eq.~(\ref{eq:nu_mfp}) because the Pauli blocking is
irrelevant near the surface where the density is small. 
The conversion rate of the neutrino energy into heat is then
\be
{dE_{\scrpt{in}}\over dt} 
= 
\kappa_\scrpt{Urca} T_c^8\, {4\pi\over 3}R_c^3 {dR\over \lambda_{e \nu}}
\ee
near the surface.\footnote{%
	Neutrino pair annihilation process also deposit
	energy in the form of lepton pairs. 
	However, the mean free path for this process may be orders 
	of magnitude larger than that of the Urca process because 
	the available transverse energy decreases as the radius increases.}
For a steady state to be maintained,
this incoming energy must be balanced by the energy going
out of the volume $dV = 4\pi R^2dR$.
We take the neutrino pair bremsstrahlung to be the dominant cooling
process of the surface due to the fact that the neutron star surface is
mainly composed of stable nuclei such as ${}^{56}_{26}$Fe.
Then heat is radiated with the rate given by \cite{shapiro}
\be
{dE_{\scrpt{out}}\over dt}
=
\kappa_\scrpt{Brem} T_s^6\, 4\pi R^2 dR\ , 
\ee
where $T_s$ is the temperature at $R$ and
\be
\kappa_\scrpt{Brem} \sim 
10^{-6} \MeV^{-5} \, \fm^{-3} \sec^{-1}
\left( {\rho_s\over \rho_\scrpt{nuc}} \right)
\;.
\ee
Equating the two, we get
\be
T_s 
\sim
0.1\,\MeV\,
\left( {T_c \over 1\,\MeV} \right)^{5/3}
Y_e^{1/6}\,
\left( {\rho_c \over \rho_\scrpt{nuc}} \right)^{1/9}
\left( {R_c \over 1\,\km} \right)^{1/2}
\left( {10\,\km \over R} \right)^{1/3}
\label{eq:TsInTc}
\;,
\ee
where $Y_e \equiv \rho_e/\rho_s$ is about $0.5$ if the Coulomb lattice
is made of ${}_{26}^{56}$Fe. 
With ${T_c\sim 5\,\MeV}$ and $R_c \sim 1\,\km$, the surface temperature of
$1\,\MeV$ is easily maintained.

\section{The Gamma-sphere and Its Interface with the Hot Neutron Star}
  
Outside the baryonic matter which makes up the neutron star, there
are high energy gammas which are escaping the surface.  Unlike the
situation for stars such as the sun, these gammas are interacting with
one another on a time scale which is small compared to the typical
expansion time for the photons, that is, the time it takes for the
local photon energy density to dilute by a factor of two.  Paczynski
and Goodman have written down hydrodynamic solutions for the evolution
of matter in this region.  
We will shortly give an argument that the baryon to photon ratio 
$N_B/N_\gamma$ is small enough for the hydrodynamic picture to be valid. 
But for now, let us assume that the baryon to photon
ratio, $N_B/N_\gamma$, is so small that one can treat the system
outside the star as an electroweak plasma.  In this case, the pressure
is
\be
    P = N_{\scrpt{dof}} {\pi^2 \over{90}} T^4
\label{eq:PInT}
\;,
\ee
where the number of degrees of freedom $N_{\scrpt{dof}}$ certainly includes
photons, may include electron-positron pairs if the temperature is
high enough, and might also include neutrinos if they were not yet
decoupled from the system.

We will assume that the time evolution of the system occurs on
time scales long compared to the time it takes light to propagate the
characteristic size scale of the system.
If this is the case, we look for a static solution to the
hydrodynamic equations,
\be
\partial_\nu T^{\nu \mu} = 0
\;.
\ee
We will ignore the effects of gravity in the gamma-sphere.  (In the
interior of the star this is not a good approximation and one must 
modify the right hand side of the above equation.)

These equations were solved by Paczynski far from the star. They are
equivalent to the algebraic equations
\begin{eqnarray}
T\gamma 
& = & 
T_s \gamma_s
\label{eq:hydro1}
\;,
\\
\noalign{\hbox{and}}
4\pi r^2 (\rho + P) v \gamma^2 
& = & 
L
\;.
\label{eq:hydro2}
\end{eqnarray}
Here the energy density of the electroweak plasma is $\rho$ and
its pressure is $P$.  This electromagnetic plasma is moving with
a velocity $v$ and Lorentz gamma factor $\gamma$.  The temperature is $T$.
The luminosity of the star is $L$.  $T_s$ is the surface temperature of
the star and $\gamma_s$ is the Lorentz gamma factor of the fluid
just outside the surface.  A little algebra shows that the minimum
velocity for the fluid which composed the gamma-sphere occurs at some
minimum radius and is $v^2 = 1/3$.  See the appendix. 
At larger distances, the velocity
of the fluid increases.  At large distances, $\gamma \sim r/R$ and
$T \sim T_s R/r$,  so the fluid rapidly become ultrarelativistic.

It is plausible that the minimum $v^2 = 1/3$ occurs at the surface
between the gamma-sphere and the neutron star.  If the neutron star is not
evaporating too rapidly, this must happen, since the rarefaction front 
associated with the electromagnetic plasma will propagate inward until it is
slowed by the surface of the star.  Moreover, if one emits free photons from
a surface at rest, the average outward velocity of the emitted photons
will be $v^2 = 1/3$.  We therefore take the minimum radius where $v^2 =
1/3$ to be the surface of the star.

The first problem which we must address in this picture is how small the
ratio $N_B/N_\gamma$ should be.  
In order that one can ignore
the effect of the baryons on the collective expansion of the fluid,
the total energy of photons at decoupling must be larger than the
total energy of the baryons
\be
N_B/N_{\gamma} < T_{\scrpt{decoupling}}/m_{\scrpt{proton}}
\; ,
\label{eq:NbOverNgamma}
\ee
where $T_{\scrpt{decoupling}}$ is the rest frame temperature 
when the photons start decouple.  

This decoupling temperature is determined by the requirement 
that the photon interaction probability beyond the decoupling radius 
(and hence the decoupling temperature) is negligible. 
The decoupling radius is in turn determined by the condition
\be
r_{\scrpt{decoupling}}
=
\lambda_\gamma 
\;,
\ee
where $\lambda_{\gamma}$ is the mean free path for photon scattering.
We expect decoupling at a temperature less than
that for which there are abundant electron positron pairs, so the
decoupling temperature is determined by photon-electron Compton scattering
\be
\lambda_{\gamma} 
=
{1\over \sigma_{\scrpt{Compton}}\rho_{\scrpt{electron}}}
\sim
{{ m_{\scrpt{electron}}^2} \over \alpha^2} {N_{\gamma}
\over N_B} T^{-3} \ ,
\ee 
where we used the fact that
$\rho_{\scrpt{electron}} = \rho_{\scrpt{proton}} =
(N_B/N_\gamma)\rho_\gamma$.
This gives for the decoupling distance
\be
R \gamma_{\scrpt{decoupling}}
\sim
r_{\scrpt{decoupling}} 
\sim
10^7\,\km\,
\sqrt{{N_B \over N_\gamma}}\,
\left({T_s\over 1\,\MeV}\right)^{3/2} 
\left({R\over 10\,\km}\right)^{3/2} \ .
\ee

The criterion that the baryons are not important (\ref{eq:NbOverNgamma})
now becomes
\be
{N_B \over N_\gamma}
<
10^{-6}
\left( {1\,\MeV \over T_s} \right)^{1/3}
\left( {10\,\km \over R} \right)^{1/3} \ .
\ee
Hence for surface temperature of 1 MeV, 
the baryon contamination must be less than one part in a million.

This small baryon contamination is hard to achieve by conventional
means.
For such an estimate for a neutron star heated from inside,
we note that on the average, an outgoing photon of energy $T$ transfers 
a momentum of order $T$ to an electron-proton pair.
Assuming a biased random walk, 
we then have too big $N_B/N_\gamma \sim 10^{-3}$ for $T \sim 1\,\MeV$.

For an estimate of the ratio for a proto-neutron star,
assume a radiation dominated atmosphere \cite{Bethe}. 
When the radiation energy density $T^4$ exceeds that of the
gravitational energy density $GM m_\scrpt{nucleon} \rho/r$, 
energy outflow in the form of baryon ejection is expected. 
This happens when 
\be
 \rho_s 
 \la 
 \rho_0 
 \equiv 
 \frac{T^4 R}{GMm_\scrpt{nucleon}} 
 \,. 
 \label{eq:rhoexpell}
\ee
If the matter and radiation flows out with the same velocity,
then the ratio $N_B/N_\gamma$ equals the ratio of the baryon density 
and the photon density.  Using the estimate 
(\ref{eq:rhoexpell}) yields again too big $N_B/N_\gamma$:
\be
{N_B\over N_\gamma}
\sim 
{\rho_0\over T^3}
\sim 
10^{-2}
\left( {R \over 10\,\km} \right)
\left( \frac{M_\odot}{M} \right)
\left( {T\over 1\,\MeV} \right)
\;.  \label{Beject}
\ee

Another estimate of the mass expulsion 
was given by Meszaros \& Rees \cite{rees}.
They assumed that the surface of the initially cold neutron star with densities
lower than $\rho_0$ is expelled.
The outer density profile of a cold neutron star is calculated from
the hydrostatic equation 
\be
  \frac{1}{\rho}\frac{dP}{dr} = -\frac{GM}{r^2} 
    \ ,\label{eom}
\ee
with the degenerate electron pressure $P\propto \rho^{4/3}$. 
They estimate a lower limit of the ejected mass ($\rho(r_0)=\rho_0$)
\be
   \Delta M\simeq 4\pi R^2 \int_{r_0}^\infty \rho(r)dr
   = 10^{-12} M_\odot \left(\frac{R}{10km}\right)^{10/3}
     \left(\frac{T}{1MeV}\right)^{16/3}
     \left(\frac{M}{M_\odot}\right)^{1/3}
   \,. \label{eq:MR}
\ee
Here they assumed a temperature of $T=63$MeV based on an estimate
of frictional
heating in mergers and consequently obtained 
$\Delta M\sim 10^{-3}M_\odot$, which is too big.
Detailed numerical calculations of mergers have
later found lower temperatures \cite{Ruffert}.  
With $T\sim 1$MeV the amount of baryon ejected is sufficiently small
according to (\ref{eq:MR}) that baryon contamination can be ignored.
However, it is only a lower estimate as can be seen by 
comparing to (\ref{Beject}).

Compared to these conventional estimates, 
one part in a million may seem unnatural. 
However there may be numerous ways to get around the above naive estimates. 
We list some of the possibilities here:
\begin{description}
\item[Strong magnetic field]\mbox{}

A surface temperature of $\sim 1\, \MeV$ can be maintained without
ejecting baryons by a magnetic field of
$B \sim 3 \times 10^{13}$\,{\rm G} \cite{murthy}. 
A magnetic field of such strength is
perhaps not unreasonable for a pulsar-like neutron star where
differential rotation or dynamo effect can generate a strong 
magnetic field. 
This strong magnetic field need not cover the whole surface of the
neutron star.   
A strong magnetic field creates a relatively baryon-free region near
the magnetic poles because charged particles follow the diverging
magnetic field lines.
These baryon-free regions near the magnetic poles would be 
effective in producing a GRB \cite{rees2}.

\item[Super heated Coulomb lattice]\mbox{}

Another possibility is the super-heated Coulomb lattice.  
Even though the Coulomb lattice near the surface of a neutron star 
should melt at temperature below $1\,\MeV$, a meta-stable state might
exist for some time before melting during which a GRB would
be produced.  
It might also be stable in a strong magnetic field.
Whether the super-heated lattice
can exist at $T \sim 1\,\MeV$ requires more careful study.

\item[Strange star]\mbox{}

When the density of matter becomes much higher than the nuclear matter
density, a strange matter may be the ground state. 

A typical strange star has a thin crust of baryons at the surface.
The fireball originating from a strange star, therefore,
will be contaminated by baryons from the crust.
Fortunately, the crust is very thin 
($M_\scrpt{crust} \sim 10^{-5}M_\odot$) \cite{cheng}
and the fireball may be relatively free of baryons.
Materials inside the crust will not be disturbed by $1\,\MeV$
temperature since
the nucleon binding energy of a strange star is a few tens of $\MeV$. 

Note that our formula for the surface temperature would not work for 
a strange star since the chemical composition of a strange star is
very much different from that of a neutron star.  However, the surface
temperature and the core temperature should still be related by a power
law.  Hence, qualitative descriptions in this and the next section
should still apply.

\end{description}

\section{Cooling of the Star}

As the star emits energy it cools.  At the temperature of order
$1\,\MeV$, the Urca process is the dominant source of the energy
loss \cite{shapiro}.
Consider a hot core of radius $R_c < R$ at $T_c$.
Then as in Eq.~(\ref{eq:CoreLoss}),
\be
{dE\over dt} = -\kappa T_c^8\, {4\pi \over 3} R_c^3 \ ,
\ee
where the minus sign indicates energy loss.

On the other hand, the total thermal energy inside the core is related to
the temperature by
\be
    E \sim N_c T_c (T_c/\mu_c) \sim  R_c^3 \rho_c \, T_c\, (T_c/\mu_c) \ , 
\ee
where $\rho_c$ is the baryon density of the core and
$\mu_c \sim \rho_c^{2/3}/m_{\scrpt{proton}}$
is the Fermi energy associated with $\rho_c$. 
We see therefore that
\be
{dT_c\over dt}
\sim  -\kappa T_c^7\, {\mu_c\over \rho_c}
\;.
\ee
This has as a solution 
\be
T_c = T_c^\scrpt{init} \left( 1 + t/t_\scrpt{scale} \right)^{-1/6}
\;,
\ee
where
\be
t_\scrpt{scale} 
\sim 
{\rho_c \over 3 \mu_c \kappa_c \left(T_c^\scrpt{init}\right)^6}
\sim
10^4\, \sec \,
\left( {1\,\MeV \over T_c^\scrpt{init}} \right)^6 
\left( {\rho_\scrpt{nuc} \over \rho_c} \right)^{1/3}
\;
\ee
is the time scale of the initial time evolution. 
For an initial temperature $T_c^\scrpt{init} \sim 10\,\MeV$, 
$t_\scrpt{scale} \sim 10^{-2}\,\sec$.
Hence, at large times $(t > 1\,\sec)$,
\be
T_c 
\sim 
\left({\rho_c \over \kappa \mu_c\, t}\right)^{1/6} 
\;.
\ee
Notice that this is independent of the initial temperature.
Since
\be
\kappa \sim 10^{-4}\,\MeV^{-7}\, \sec^{-1}\, \fm^{-3}
\left( {\rho_c\over \rho_{\scrpt{nuc}}} \right)^{2/3}\ ,
\ee
we then have
\be
T_c \sim 
5\,\MeV \,
\left({1\,\sec \over t}\right)^{1/6}
\left({\rho_{\scrpt{nuc}} \over \rho_c}\right)^{1/18}
\;.
\ee
The natural time scale of a second for the time evolution of the
burst is a simple consequence of neutrino radiation from a neutron star size 
object at nuclear matter scale density.

We can now use Eq.~(\ref{eq:TsInTc})
to calculate the evolution of the surface temperature. 
The result is
\be
\left({T_s\over 1\,\MeV}\right)
& \sim &
\left({1\,\sec\over t}\right)^{5/18}
Y_e^{1/6}\,
\left( {\rho_c \over \rho_\scrpt{nuc}} \right)^{1/54}
\left( {R_c \over 1\,\km} \right)^{1/2}
\left( {10\,\km \over R} \right)^{1/3}\ .
\ee
This evolution of the temperature has many consequences.  First there
should
be a correlation between the arrival time and the energy of photons from
the
gamma ray burst.  Second, the luminosity should go like 
the typical energy to the power of $\sim 1.1$ at any time.  
It seems the data is in rough agreement with such a prediction.  

Another consequence is a hardening of the spectrum of emitted photons
at lower energy.  If the star had only a fixed temperature, then the
differential photon spectrum would be
\be
     {{dN} \over {d\omega}} \sim {{\omega^2} \over {e^{\omega/T_s} -1}}\ ,
\ee 
which for $\omega \ll T_s$ scales like $\omega$.  If on the other
hand, we emit from a range of temperatures, the distribution will be
hardened at the low frequency end so long as 
$\omega > T_{\scrpt{min}}$, where $T_{\scrpt{min}}$ is the minimum 
temperature for which there is thermal emission.
We estimate the rate as $dN \sim E^3 dt \sim T_s^3 dt$ 
or
\be
{dN\over d\omega} \sim \omega^{-8/5}
\ee
since $T_s \sim 1/t^{5/18}$.
The spectrum is therefore a factor of $\omega^{-13/5}$ harder at low energies
than is predicted by the black body law.  This seems to work in the
correct general direction as the data.  A representative of a data
parameterized by $\exp(-\omega/T_0)/\omega$ with $T_0 = 0.505\MeV$
\cite{murthy} and 
our result $\omega^{-8/5}$ is plotted in Fig.~1.

\bigskip
\bigskip

\begin{figure}[htb]
\begin{center}
\leavevmode
\epsfxsize=12cm
\epsfbox{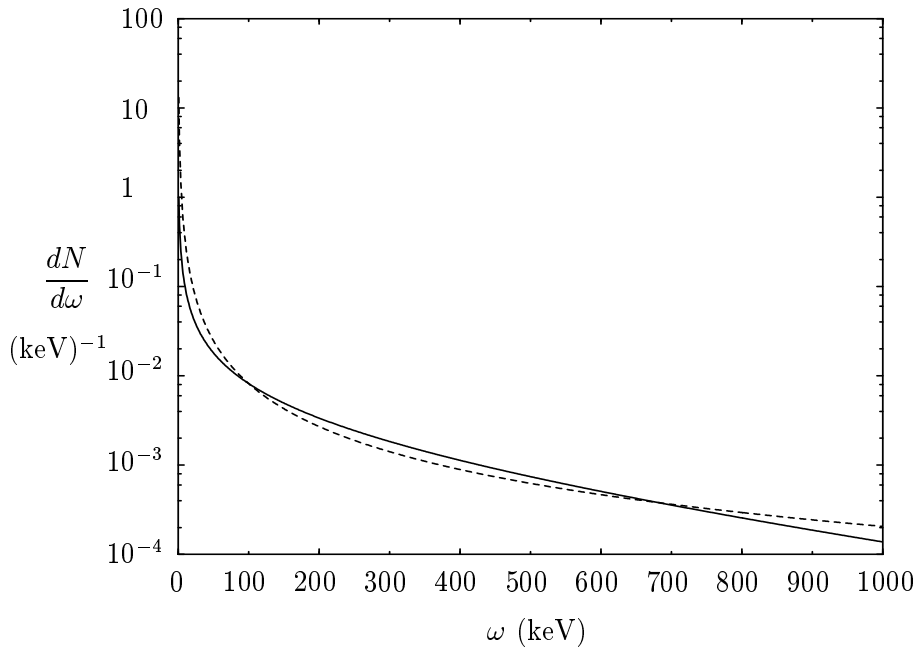}
\end{center}
\caption{
Photon energy spectrum.  The solid line represents 
a parametrization of an observed spectrum,
$dN/d\omega \sim \exp(-\omega/T_0)/\omega$ with $T_0 = 0.505\MeV$, and
the broken line represents our result $dN/d\omega \sim \omega^{-8/5}$.
}
\end{figure}

\section{Summary and Conclusions}
We have attempted to model gamma ray bursts as emission from hot
neutron stars with a surface temperature
$T_s \sim 1\,\MeV$.  In the crude order of magnitude
estimates we have made, we see that we can describe the gross features 
of such emission.  We get a qualitatively valid description of the time
evolution.  
We also argued that
a small ratio of $N_B/N_{\gamma}$ in the hot radiating 
gamma-sphere of the star can be achieved via strong magnetic field or a
super-heated Coulomb lattice.
Whether or not this model will work in detail needs further analysis.  

In this paper we have tried to ignore the detailed dynamics
which will describe the region which we refer to as the
hot neutron star.  If the description above satisfies experimental 
constraints, then one can proceed with confidence to a more detailed
description.

\section*{Acknowledgments}
L. McLerran would like to thank M. Voloshin for numerous helpful
discussions and encouragement.   Research supported by the U.S.
Department of Energy under grants
No. DOE Nuclear DE--FG06--90ER--40561 and  DE-FG02-87ER40328. 

\appendix

\section{A solution to Eqs.~(\protect{\ref{eq:hydro1}}) and
(\protect{\ref{eq:hydro2}})}

Using $P = a T^4/3$ and the ultra-relativistic approximation 
$\rho = 3P$, one can easily obtain the following solution for $v$
\be
v(r) = {2\over \sqrt{3}}\cos(\theta(r)/3) 
\ee
where
\be
\theta(r) 
= 
\pi - \tan^{-1}\left(\sqrt{ {r^4\over r_{\scrpt{min}}^4} - 1 }\right)
\ee
Here, the minimum radius is defined to be
\be
r_{\scrpt{min}} 
\equiv
\sqrt{9\sqrt{3} L\over 32 a T_s^4 \pi}
\ee
When $r = r_{\scrpt{min}}$, $v^2(r_{\scrpt{min}}) = 1/3$.
Note that this $r_{\scrpt{min}}$ is different from $r_0$ defined in 
\cite{paczynski}.  In Ref.\cite{paczynski}, $r_0$ defines the radius
where $\gamma = 1$ in the large $r$ approximation.  However, this is a
definition of convenience.  As is clear from our solution, the minimum
speed is that of the sound speed of a photon gas.  Hence, $\gamma$ never
goes to $1$.

\end{document}